\documentclass[12pt,thmsa]{article}

\usepackage{cite}

\begin{document}

\title{Nonsensical models for quantum dots}
\author{Francisco M. Fern\'{a}ndez \\
INIFTA (UNLP, CCT La Plata--CONICET), \\
Divisi\'{o}n Qu\'{i}mica Te\'{o}rica, Sucursal 4, Casilla de Correo 16, \\
1900 La Plata, Argentina}
\maketitle

\begin{abstract}
We analyze a model proposed recently for the calculation of the energy of an
exciton in a quantum dot and show that the authors made a serious mistake in
the solution to the Schr\"{o}dinger equation.
\end{abstract}

Not long ago Hassanabadi and Rajabi\cite{HR09} proposed a simple model for a
spherical quantum dot. They treated the nonrelativistic Schr\"{o}dinger
equation with a phenomenological potential as a quasi--exactly solvable
problem so that the model parameters had to depend on the given state of the
system. Although they imported particle masses, excitation energies and
other model parameters from experimental data, they never contrasted their
model outputs with actual experiments. More precisely, they calculated model
properties and never compared them with independent physical data although
they stated mischievously otherwise. Worst of all, they wrongly calculated
the excitation energy of the quantum dot by subtracting the ground state of
one model from the excited state of a quite different one. Their mistake was
based on the fact that their model parameters depended on the state of the
problem as we already disclosed in a later comment\cite{F09}.

Recently, Hassanabadi and Zarrinkamar\cite{HZ09} came back with a new model
for the exciton in a quantum dot that we analyze in what follows.

The authors chose the nonrelativistic Hamiltonian
\begin{equation}
\hat{H}=\sum_{i=e,h}\left( \frac{\mathbf{p}_{i}^{2}}{2m_{i}}+\frac{1}{2}%
m_{i}\omega _{0}^{2}r_{i}^{2}\right) -\frac{e^{2}}{\varepsilon |\mathbf{r}%
_{e}-\mathbf{r}_{h}|}  \label{eq:H_model}
\end{equation}
where the subscripts $e$ and $h$ stand for the electron and the hole,
respectively, and the motion is in the plane\cite{HZ09}.

The Schr\"{o}dinger equation for this problem is separable in the center of
mass $\mathbf{R}=(m_{e}\mathbf{r}_{e}+m_{h}\mathbf{r}_{h})/(m_{e}+m_{h})$
and relative $\mathbf{r}=\mathbf{r}_{e}-\mathbf{r}_{h}$ coordinates and we
are left with the eigenvalue equations
\begin{eqnarray}
\left( \frac{\mathbf{p}_{R}^{2}}{2M}+\frac{1}{2}M\omega _{0}^{2}R^{2}\right)
Q &=&E_{R}Q  \nonumber \\
\left( \frac{\mathbf{p}_{r}^{2}}{2\mu }+\frac{1}{2}\mu \omega _{0}^{2}r^{2}-%
\frac{e^{2}}{\varepsilon r}\right) \psi &=&E_{r}\psi  \label{eq:H_R,H_r}
\end{eqnarray}
where $\mu =m_{e}m_{h}/M$ and $M=m_{e}+m_{h}$ are the reduced and total
masses, respectively. Notice the misprint in their equation (6). The first
equation is just a harmonic oscillator that we do not discuss any further
and the second one is separable in spherical coordinates $\psi (\mathbf{r)}%
=\Phi (r)e^{im\phi }$, where $m=0,\pm 1,\ldots $ is the angular quantum
number. On comparing their equations (8) and (9) one has the impression that
the authors carried out this separation without being clearly aware of it%
\cite{HZ09}.

The radial part is a solution to\cite{HZ09}
\begin{equation}
\Phi ^{\prime \prime }+\frac{1}{r}\Phi ^{\prime }-\frac{m^{2}}{r^{2}}\Phi
+\alpha \Phi -\beta r^{2}\Phi +\frac{\gamma }{r}\Phi =0  \label{eq:difeq_Phi}
\end{equation}
where $\alpha =2\mu E_{r}/\hbar ^{2}$, $\beta =\mu ^{2}\omega _{0}^{2}/\hbar
^{2}$ and $\gamma =2\mu e^{2}/(\hbar ^{2}\varepsilon )$. This equation is
not exactly solvable as everybody (except the authors) knows.

Hassanabadi and Zarrinkamar\cite{HZ09} proposed the ansatz
\begin{equation}
\Phi (r)=r^{m}e^{-\sqrt{\beta }r^{2}/2}\sum_{n=0}^{\infty }a_{n}r^{n}
\label{eq:Phi_ansatz}
\end{equation}
and derived a recurrence relation for the coefficients that should read
\begin{eqnarray}
a_{0} &=&1  \nonumber \\
a_{1} &=&-\frac{\gamma }{2m+1}  \nonumber \\
a_{n+2} &=&\frac{\left[ 2\sqrt{\beta }(m+n+1)-\alpha \right] a_{n}-\gamma
a_{n+1}}{(n+2)(2m+n+2)}  \label{eq:rec_an}
\end{eqnarray}
Notice that there is a misprint in the denominator of their equation (17c).

At this point the authors treated this problem as if it were the textbook
harmonic oscillator or hydrogen atom and based on the argument that ``the
series must be bounded for $n=n_{r}$'' they chose
\begin{equation}
\alpha =2\sqrt{\beta }(m+n_{r}+1)-\gamma \frac{a_{n_{r}+1}}{a_{n_{r}}}
\label{eq:alpha}
\end{equation}
and supposedly obtained the exact energies and wavefunctions
\begin{equation}
E_{0,m}=2\hbar \omega _{0}(m+1)+\frac{4\mu e^{4}}{\varepsilon \hbar
^{2}(2m+1)}  \label{eq:E_0m}
\end{equation}
\begin{equation}
\Phi _{0,m}(r)=N_{0m}r^{m}e^{-\sqrt{\beta }r^{2}/2}\left( 1-\frac{\gamma r}{%
2m+1}\right)  \label{eq:Phi_0m}
\end{equation}
respectively, for $n_{r}=0$. However, the reader may verify that $\Phi
_{0,m}(r)$ is not a solution to Eq.~(\ref{eq:difeq_Phi}) unless the
additional condition
\begin{equation}
2\sqrt{\beta }(2m+1)-\gamma ^{2}=0  \label{eq:beta(gamma)}
\end{equation}
is satisfied.

As we said above, this model is not exactly solvable. However, one can
obtain solutions for particular values of the model parameters with the well
known consequence that they have to depend on the quantum numbers of the
chosen state. We already discussed this feature of quasi--exactly solvable
problems in our earlier criticism of the Hassanabadi and Rajabi's paper\cite
{F09} but the former author did not appear to take notice. Under such
conditions the energy should be
\begin{equation}
E_{0,m}=\hbar \omega _{0}(m+2)  \label{eq:E_0m(correct)}
\end{equation}

We can easily trace the mistake in Hassanabadi and Zarrinkamar's reasoning
to the fact that Eq.~(\ref{eq:rec_an}) is a three--term recurrence relation.
Therefore, the condition $a_{n_{r}+2}=0$ alone is insufficient to force $%
a_{j}=0$ for $j>n_{r}+2$ as in the well known cases of the pure harmonic or
Coulomb interactions. In order to have an exact solution we should set the
model parameters so that $a_{n_{r}+2}=0$ and $a_{n_{r}+3}=0$ which accounts
for the additional condition (\ref{eq:beta(gamma)}) in the particular case $%
n_{r}=0$.

In addition to what has been said, notice that the wavefunction (\ref
{eq:Phi_0m}) with $m=0$%
\begin{equation}
\Phi _{0,0}(r)=N_{0}e^{-\sqrt{\beta }r^{2}/2}\left( 1-\gamma r\right)
\label{eq:Phi_00}
\end{equation}
exhibits a node, as shown in their Fig.~2\cite{HZ09}, and therefore it
cannot be part of a ground state. In other words, the authors failed to
obtain the ground--state wavefunction of their model.

We conclude that all the results obtained by Hassanabadi and Zarrinkamar\cite
{HZ09} from ``the elegant idea of series method'' are based on wrong
equations and therefore completely useless. Even if they had not made the
mistakes just indicated their results would have been utterly useless
because they never tested the output of their model. They were satisfied
with the calculation of an ``exciton energy for various states'' that are
mere numbers without any meaning.

It is also curious that the authors stated that ``the results are comparable
to those of variational exact diagonalization, full configuration
interaction, Hartree--Fock and 1/N methods''. However, they did not provide
any reference or comparison to support their claims. If the authors had
tried any such comparison they would have probably realized their mistake.
Besides, it is not clear to us what they meant for full configuration
interaction and Hartree--Fock with respect to their fully separable model.

The authors also claimed that their approach ``could be fitted for any
desired material'' meaning that they input some experimental data and output
results that they never compared with independent data. Nowadays, this
strategy seems to be sufficient for publishing a paper in certain journals%
\cite{HR09,HZ09}.

The reader may wonder why a supposedly respectable journal published such a
paper. Well, that journal has a long history of publishing wrong,
nonsensical and ridiculous papers as we have already denounced in earlier
articles\cite{F07, F08b, F08c, F08d, F08e, F08f, F09d, F09e}. You may find,
for example, solutions to the linear and nonlinear Schr\"{o}dinger equation
that are not square integrable, power series expansion of well known
functions that are unphysical solutions to physical or unphysical equations,
useless power series approaches to models for nonlinear dynamics, and many
other such horrific examples\cite{F07, F08b, F08c, F08d, F08e, F08f, F09d,
F09e}. The editors and referees of that journal are satisfied, for example,
with a model for a prey--predator system that predicts a negative number of
rabbits\cite{F08d}.

\end{document}